\documentstyle[preprint,aps]{revtex}

\tighten
\begin{document}

\title{ Comparison of techniques for computing shell-model
effective operators  }
\medskip

\author{Michael Thoresen,
        Petr Navr\'atil\footnote{On leave of absence from the
             Institute of Nuclear Physics,
             Academy of Sciences of the Czech Republic,
             250 68 \v{R}e\v{z} near Prague, Czech Republic.},
        and Bruce R. Barrett }

\medskip

\address{          Department of Physics,
                   University of Arizona,
                   Tucson, Arizona 85721 }

\maketitle

\bigskip

\begin{abstract}
   Different techniques for calculating effective operators
within the framework of the shell model using the same 
effective interaction and the same excitation spaces 
are presented. Starting with the large-basis no-core approach,
we compare the time-honored perturbation-expansion approach 
and a model-space truncation approach. Results for the electric 
quadrupole and magnetic dipole operators are presented for $^6$Li.
The convergence trends and dependence of the effective operators 
on differing excitation spaces and Pauli $Q$-operators is studied.
In addition, the dependence of the electric-quadrupole effective charge 
on the harmonic-oscillator frequency and the mass number, for $A=5,6$, 
is investigated in the model-space truncation approach.
\end{abstract}

\bigskip
\bigskip
\bigskip

\narrowtext

\section{ INTRODUCTION }

While considerable effort has been devoted to derive the
effective interaction used in the shell-model calculations
from the nucleon-nucleon interaction \cite{B67,HJKO95,EO77},
less work has been done to understand the effective operators 
employed in calculating different nuclear, 
usually electromagnetic, properties. 
A microscopic derivation of effective operators has been
only partially successful. It is well known that effective
proton and neutron charges must be employed to describe
the E2 transitions and moments. These charges are quite
different from the free nucleon charges, typically the values of
$e_{\rm eff}^{\rm p}\approx 1.5 e$ and 
$e_{\rm eff}^{\rm n}\approx 0.5 e$
are obtained for both light and heavy nuclei.
It should be noted that these effective charges correspond
to a severe truncation to a single-major-shell, 
or $0\hbar\Omega$, space.
Attempts to derive these charges microscopically, usually by
perturbation theory \cite{EO77}, or by an ``expanded shell-model''
approach \cite{FP78}, yielded much smaller values.
We note that in the ``expanded shell-model'' approach, typically results
of a $(0+2)\hbar\Omega$ calculation were truncated to the $0\hbar\Omega$
space to derive the effective charges.

  In a previous paper we described a truncation procedure for
determining effective charges in small shell model spaces using the
results of large-space no-core calculations \cite{NTB97}. 
In that study, we used a $6\hbar\Omega$ model space for $^6$Li,
i.e., including excited states up to $6\hbar\Omega$ above the
unperturbed ground-state configuration.
This procedure yielded $0\hbar\Omega$ effective charges 
consistent with empirical values. 
Since previous microscopic
perturbation-theory calculations have typically produced effective
charges, which are too small compared with the empirical values, it
is of interest to compare perturbation-theory calculations and
calculations in a truncated model-space in an attempt to understand the
reason for these differences.

  To carry out this study we propose to perform three calculations:
(1) no-core, (2) perturbation theory,  and (3) model-space truncation,
all using the same
nucleon-nucleon potential and assumptions for determining the nuclear
reaction matrix $G$. In addition, to shed more light on the 
harmonic-oscillator-frequency and the mass-number dependence, we 
supplement the $^6$Li results of Ref. \cite{NTB97} with calculations
for other $\hbar\Omega$ values and also present new results for $A=5$
system. 

  In section II we describe the formalism used in performing the
three calculations. The no-core results are presented 
in section III. The perturbation calculation results 
and the model-space truncation results 
are discussed in section IV and V, respectively. 
Conclusions are given in section VI.

\section{ FORMALISM }
All three approaches used in the study begin with
the calculation of the  Bruckner reaction matrix (or $G$-matrix)
defined as
\begin{equation}
\label{Gmatrix}
G(\varepsilon)=V+V{Q^{P}\over{\varepsilon-H_{0}}}G(\varepsilon).
\end{equation}
Either the starting energy $\varepsilon$ is parametrized by 
\begin{equation}\label{omega}
\varepsilon = \epsilon_a + \epsilon_b + \Delta,
\end{equation}
with $\Delta$ being independent of $\epsilon_a$ and $\epsilon_b$, or
the dependence of the $G$-matrix on $\varepsilon$ is
removed by taking into account the folded diagrams by means of the
Lee-Suzuki \cite{LS80} approach following the procedure described in 
Ref. \cite{NB96}.
In equation (\ref{Gmatrix}) $V$ is the free nucleon-nucleon (N-N) interaction.
The starting energy
$\varepsilon$ represents the initial energy of the two nucleons in the
medium.
$H_{0}$ is the unperturbed Hamiltonian of the system, and
$Q^{P}$ is the Pauli projection operator, which projects onto
two-particle states that are not already occupied. 
The $G$-matrix
is evaluated using the method of Barrett, Hewitt and McCarthy
\cite{BHM71} which is similar to the reference-spectrum method
\cite{BBP63}, in that a reference matrix $G^{R}$ is calculated and
the reaction matrix, or $G$-matrix, is then obtained by matrix inversion.
The computation of the $G$-matrix employed in our calculations 
uses the Nijmegen II or the Reid93
N-N interaction \cite{SKTdS94} in a two-particle 
$Q^{P}$-space shown in Fig. \ref{wingsfig} and defined such that 
\begin{eqnarray}
Q^{P} & = \; 0 &\;{\rm \;for}\;(N_{1}+N_{2})\leq N_{\rm max},\;N_{1}=0,
\;{\rm or}\;N_{2} = 0 \; , \\
\nonumber Q^{P} & = \; 1 &\;{\rm \;for\;all\;other\;} 
N_{1} {\rm\; and\;} N_{2} \; , \\
\nonumber {\rm with} & & N_{1} = 2n_{1} + l_{1} \; 
{\rm and} \; N_{2} = 2n_{2} + l_{2}\;.
\end{eqnarray}
In most calculations we use $N_{\rm max}=8$.
Setting $Q^{P} = 0$ for $N_{1}+N_{2}\leq 8$ forces the intermediate 
excitations
to have an energy greater than 8$\hbar\Omega$. We will refer to the 
$G$-matrix calculated with a $Q^{P}$-space defined this way as an 
8$\hbar\Omega$ $G$-matrix. Setting $Q^{P} = 0$ for $N_{1}=0$ or 
$N_{2} = 0$
prevents us from scattering into intermediate states that are in the
$0s_{1/2}$ state, which is already fully occupied. Hence, the Pauli Exclusion
Principle prevents us from scattering into these states. These portions
of the $Q^{P}$-space are referred to as the ``wings" of the $Q^{P}$-projection
operator. In calculations which do not assume there is a fully occupied
$^{4}$He core, these $0s_{1/2}$ states may be accessible and the wings 
may not be needed in the calculation of the $G$-matrix. For calculations
without wings we use the $Q^{P}$-operator of Fig. 
\ref{wingsfig}(b) defined such that 
\begin{equation}
Q^{P}  = \; 0 \; {\rm \;for}\;(N_{1}+N_{2})\leq N_{\rm max}, 
\end{equation}
In theory the wings should extend to infinity, but computational
limits necessitate putting a finite limit on the extent of the wings.
Barrett, Hewitt and McCarthy \cite{BHM71} investigated truncating
the wings at different values and found that there is little effect 
in extending the wings past $N=10$. Consequently, in our study we 
have truncated the wings at $N=10$. Because we obtain slightly different
$G$-matrix elements, when we use different $Q^{P}$-operators, we
will study the effect that including wings has on the calculations.
For the purpose of comparison,
both the no-core and the perturbation-expansion calculations are done
with and without wings.

\subsection{ NO-CORE APPROACH }

  The no-core approach refers to large-basis shell-model calculations
performed in a model space of several major harmonic-oscillator
shells. In this approach all $A$
nucleons of a given nucleus are active for a complete $N\hbar\Omega$
basis space with a large value for $N$ \cite{ZBVHS95}. 
(An $N\hbar\Omega$ basis space is one which includes all allowed
configurations
up to an energy of $N\hbar\Omega$ above the unperturbed ground state.)
Due to the no-core assumption, the effective interaction used in 
the calculations is simplified as no hole states are present.
In the approach taken, the effective, in general $A$-body, interaction 
is determined for 
a system of two nucleons only and subsequently used in many-nucleon
calculations. As discussed in the beginning of this section,
either the two-nucleon $G$-matrix with a particular parametrization 
of the starting energy or the two-nucleon $G$-matrix with
the folded diagrams taken into account by means of the Lee-Suzuki 
approach, is employed as the two-body effective interaction
in our no-core shell-model calculations.
By working in a complete $N\hbar\Omega$ basis space
with a single-particle harmonic-oscillator
Hamiltonian as our unperturbed Hamiltonian, we can guarantee that all
excluded configurations involve an energy of at least
$(N+2)\hbar\Omega$, which should limit any intruder-states difficulties
to the less interesting physical states higher in the spectrum
\cite{SW72}. That is, the larger the value of $N$, the
better the guarantee that we have included the major configurations
making up the physical low-lying states.
Using complete $N\hbar\Omega$ harmonic-oscillator
spaces allows us to project out the spurious center-of-mass
components in the wave functions \cite{vHG83,BG77,BL61}.
Note, however, that the calculations using the $G$-matrix with the wings
mix center-of-mass and relative-coordinate configurations,
as there is no orthogonal transformation between the two-particle states
and the relative and center-of-mass coordinate states of the two 
interacting particles in this case. We checked the effect of mixing of the 
center-of-mass and relative-coordinate configurations in the 
$6\hbar\Omega$ calculation by varying the projection parameter
used in the Hamiltonian and found that it is a very small effect.    

In the no-core shell-model calculations presented here we calculate
eigenenergies, electromagnetic properties using the bare nucleon charges,
and the point-nucleon radii of $^6$Li. In addition, 
we use the no-core shell-model calculation results for
$^5$He and $^5$Li in section V.

  In order to gain insight into the model-space dependence of the
no-core calculation, three different model spaces are investigated.
For these three model spaces we use $G$-matrix elements evaluated in
such a manner as to include all two-particle states with unperturbed 
energies up to $4\hbar\Omega$, $6\hbar\Omega$, and $8\hbar\Omega$
relative to the harmonic-oscillator ground state,
corresponding to
excitations of $2\hbar\Omega$, $4\hbar\Omega$, and $6\hbar\Omega$
above the lowest-energy configuration of $^{6}$Li, respectively.

\subsection{ PERTURBATION EXPANSION }

  Once we have determined the $G$-matrix we may employ a perturbation
expansion for effective operators calculated to second order to 
determine effective charges.
The perturbation-expansion diagrams for effective operators 
included in our study follow the
work of Siegel and Zamick \cite{SZ70}. Zeroth-order terms for effective
charges are simply matrix elements of 
the bare operators. The two first-order terms, including
their exchange diagrams, contain one
intermediate state and have one interaction between the valence
particles and the core particles. These zeroth-order and first-order
diagrams are shown in Fig. \ref{1bop} (a-c). The two first-order diagrams
are evaluated as
\begin{eqnarray}
\langle b |[O^{L\tau} a]^{b} \rangle
& = & \sum_{p,h} (1+P_{ph}) (-1)^{(L+\tau+j_{p}+j_{h}+1)} 
\frac{1} {\epsilon_{b}-(\epsilon_{a}+\epsilon_{p}-\epsilon_{h})} 
\nonumber \\
 & & \times \sum_{J,T} (2J+1) (2T+1) 
\sqrt{\frac{(2j_{p}+1)}{(2j_{b}+1)}} \;
\langle (j_{b}\;j_{h})^{JT}|G|(j_{a}\;j_{p})^{JT}\rangle \nonumber \\
 & & \times W \left(L, j_{p},j_{b},J; j_{h}, j_{a} \right)
W \left(\tau,\frac{1}{2},\frac{1}{2},T;\frac{1}{2},\frac{1}{2}\right)
\langle p |[O^{L\tau} h]^{p} \rangle \; . 
\label{effop1}
\end{eqnarray}
The operator $P_{ph}$ exchanges the labels $p$ and $h$, 
except for the energy denominator where its action results in 
the exchange of $a$ and $b$. It
changes diagram \ref{1bop}(b) into diagram \ref{1bop}(c). 
The quantities W are Racah coefficients and the reduced matrix
element convention of Mavromatis, Zamick and Brown \cite{MZB66}
is employed, where the reduced matrix element
$\langle p |[O^{L\tau} h]^{p} \rangle$ is defined by
\begin{equation}
\langle p|O^{L\tau}_{M}|h\rangle = (LMj_{h}m_{h}|j_{p}m_{p})
\langle p |[O^{L\tau} h]^{p} \rangle \;.
\label{reduced}
\end{equation}
Instead of using approximate energy denominators set equal to multiples of
$\hbar\Omega$, as done by Siegel and Zamick, we utilize energy denominators
determined from calculated single-particle energies (e.g., $\epsilon_{a}$) 
obtained
using a second-order perturbation expansion that involves evaluating
the three one-body diagrams, shown in Fig. \ref{spe2nd}. The formulas for
these three diagrams are given in our study of effective interactions
\cite{TZB96}.
Using calculated single-particle energies will give a more
accurate description of the differences in the energies of the
particles involved in determining the energy denominators.
All configurations up to $6\hbar\Omega$ above the ground state are
included in the calculation of the energy denominators.
The same single-particle energies were employed in
all calculations, regardless of the size of the intermediate excitation
space being used for the effective-operator diagrams.

  Second-order terms for effective
operators include all terms with two intermediate states and have two
sets of interactions involving the core and excited particles. Some 
examples of the second-order diagrams for the effective charge
are shown in Fig. \ref{1bop} (d-l). 
Diagram (d) is an example
of a second-order diagram obtained by inserting a $G$-interaction into
a first-order diagram, in this case diagram (b).
Figures \ref{1bop}(e) and (f) are examples of second-order 
diagrams, which have no counterpart in first-order.
We do not include wave function renormalizations 
in the calculation.

  In calculating the effective charges, the result for
each diagram was divided by the zeroth-order diagram for
normalization, thus the renormalized zeroth-order term is identically
one for both the T=0 and T=1 cases, i.e., the isoscalar and isovector terms,
respectively. The effective proton charge $e^{\rm p}_{\rm eff}$ and the
effective neutron charge $e^{\rm n}_{\rm eff}$ are given in terms 
of the isoscalar 
effective charge $e^{0}_{\rm eff}$, evaluated with T=0, and the isovector 
effective charge $e^{1}_{\rm eff}$, evaluated with T=1,
according to the following equations
\begin{equation}
e^{\rm p}_{\rm eff} = {\textstyle{\frac{1}{2}}} 
(e^{0}_{\rm eff} + e^{1}_{\rm eff}),
\label{peffch}
\end{equation}
\begin{equation}
e^{\rm n}_{\rm eff} = {\textstyle{\frac{1}{2}}} 
(e^{0}_{\rm eff} - e^{1}_{\rm eff}),
\label{neffch}
\end{equation}
where $e^{0}_{\rm eff}$ and $e^{1}_{\rm eff}$ are determined 
by evaluating the different diagrams to various orders in 
the perturbation expansion (See Ref. \cite{SZ70}, for example).

\subsection{ MODEL-SPACE TRUNCATION }

To make a direct comparison with the perturbation results we apply
the model-space truncation formalism \cite{NTB97}
to the no-core $6\hbar\Omega$ (or $4\hbar\Omega$) calculation 
in order to derive an equivalent description in the $0\hbar\Omega$ 
space. 
That is, we take the results of the large-space $6\hbar\Omega$
calculation and truncate (i.e., project) them into the
$0\hbar\Omega$ space, so as 
to construct an effective $0\hbar\Omega$ Hamiltonian.
The Lee-Suzuki starting-energy independent similarity transformation
method \cite{LS80} is used, which gives the effective Hamiltonian 
$PH_{\rm eff}P = PHP + PHQ\omega P$,
with the transformation operator $\omega$ 
satisfying $\omega=Q\omega P$. This operator is obtained 
from the large-space $0\hbar\Omega$ dominated
eigenstates using the relations
\begin{eqnarray}\label{eisigma}
\langle\alpha_Q|k\rangle &=&\sum_{\alpha_P}
\langle\alpha_Q|\omega|\alpha_P\rangle \langle\alpha_P|k\rangle 
\; , \\
\langle\alpha_Q|\omega|\alpha_P\rangle &=& \sum_{k\in{\cal K}}
\langle\alpha_Q|k\rangle
\left[ \langle\alpha_P|k\rangle \right]^{-1} \; .
\end{eqnarray}
The states $|\alpha_Q\rangle$ and $|\alpha_P\rangle$ are the Q 
space and P space basis states, respectively, and $\cal K$ is 
the set of eigenstates $|k\rangle$ that we wish to reproduce 
in the truncated model space. Note that here the $P$ space is the 
$0\hbar\Omega$ model space.
Using this operator, the effective hamiltonian can be constructed
(see Ref.  \cite{NB96}), 
and a general effective operator is then obtained as 
\cite{NGK93,KEHLSO93,SO95}
\begin{equation}\label{hermeffop}
\bar{O}_{\rm eff} =
\left[P(1+\omega^\dagger\omega)P\right]^{-\frac{1}{2}}
(P+P\omega^\dagger Q)\hat{O}(P+Q\omega P)
\left[P(1+\omega^\dagger\omega)P\right]^{-\frac{1}{2}}
\; .
\end{equation}
In the particular application of this formalism to $^6$Li,
with $\hat{O}$ being one-body electromagnetic operators,
we obtain, in general, two-body effective operators. These can be then 
separated into a one-body part with the help of effective 
charges and a two-body part \cite{NTB97}.

\section{NO-CORE CALCULATION RESULTS}

  The results using the large-model-space no-core calculations, 
with the $G$-matrix derived from the Nijmegen II potential,
show good agreement with $^{6}$Li experimental values, as shown
in Table \ref{nocoropwing}. With the harmonic-oscillator parameter 
$\hbar\Omega$ = 14 MeV, and
the $\Delta$ of equation (\ref{omega}) chosen as -25 MeV to reproduce 
the experimental binding energy of $^{6}$Li, the electric quadrupole
moment, magnetic dipole moment and root-mean-square proton radius,
calculated in the 6$\hbar\Omega$ model space, are
all within about 10\% of their respective experimental values.

  From Table \ref{nocoropwing} we can see a very small dependence
of the magnetic dipole moment of the ground state of $^{6}$Li
on the model-space size. Each successive increase in the size of the
model space results in a smaller increase in the calculated value for the
magnetic dipole moment. The calculated value of the magnetic dipole
moment gets closer to the experimental value as the size of the 
model space increases.

  The calculated root-mean-square proton radius is somewhat smaller
than the experimentally determined value. 
There is a slight model-space dependence for the calculated value, where 
the calculated root-mean-square proton radius tends to increase 
as the size of the model space increases. 
This model-space dependence is
too small to account for the discrepancy between the calculated and
the experimentally determined values, the calculated value being about
15\% smaller than the experimental value. This is a long-standing
problem, where calculations which reproduce the correct binding energy
tend to give a radius that is too small and calculations, which produce
the correct radius, tend to underbind the system \cite{Neg82}. Since we
have chosen to reproduce the correct binding energy by our choice of
$\Delta$ = -25 MeV in the starting energy, we can expect to calculate a
radius that is too small.

  No-core calculations of the electric quadrupole moment for the ground state
of $^{6}$Li closely match the small, negative experimental value of
-0.082. Calculations performed in the 2$\hbar\Omega$ model space give us a
small, negative value that is on the order of the experimental value.
Going to larger model spaces improves the result giving us a value
that is very close to the experimental value for the electric
quadrupole moment of $^{6}$Li. The trend of convergence of the
electric quadrupole moment with larger model spaces is not so
consistent as the trend observed in the calculation of the
magnetic dipole moment, discussed previously. 

  The difference between the results obtained in the no-core approach
for the ground-state EM moments with and without wings
depends upon the operator studied. For the
magnetic dipole operator, the $0s_{1/2}$ state cannot be connected with 
the higher-lying states contained in the wings. Thus, there is no direct
contribution to the magnetic dipole operator from the wing portion
of the $Q^{P}$-operator and the effect
of the wings upon the calculation of the magnetic dipole moment is
essentially nonexistent. 
The root-mean-square proton radius exhibits a noticeable
difference, when calculated with and without wings. In each
of the model spaces studied the calculation with wings results in a
smaller binding energy, increasing the radius. The wings also include 
states with a greater portion of their wave function at large radii. The
combination of these two factors leads us to expect that calculations
with wings will result in a greater calculated root-mean-square proton
radius, which we observe in Table \ref{nocoropwing}. 
In all model spaces studied, including the wings in the
calculation of the $G$-matrix causes the no-core calculated value of
the electric quadrupole moment to be smaller than the calculation
without the wings.
The calculations for the electric quadrupole operator with and
without wings grow closer together as we increase the size of the model
space. The strong $r^{2}$-dependence and the contributions from the
expanded portions of the model space cause the electric quadrupole moment 
to show the greatest dependence upon the wings of the three operators 
calculated.
The number of
states included in the wings decreases as we go to larger model spaces,
so we would expect the difference  
between the calculations with and without wings to diminish in the
larger model space calculations. The diminishing effect of the wings
can be clearly seen in Table \ref{nocoropwing}.

\section{PERTURBATION CALCULATION RESULTS}

  We now turn to our study of the effective operators and 
effective charges obtained in the perturbation-expansion
method. We utilize the same $G$-matrix as that used for the $6\hbar\Omega$
$^6$Li no-core calculations discussed above.
In evaluating the various diagrams,
single-particle energies, calculated to second-order in
perturbation expansion, were employed to determine the energy denominators.
These single-particle energies were calculated using
\begin{equation}
\label{speeps}
\epsilon_{i} = t_{i} + u_{i} \;,
\end{equation}
where the $u_{i}$ was determined by evaluating the first- and second-order 
diagrams shown earlier in Fig. \ref{spe2nd}.
For the purpose of comparison with calculations done by others 
\cite{SZ70,ES71},
all diagrams have also been evaluated using
multiples of $\hbar\Omega$ for the energy denominators. 
The energy spacing between major shells using calculated single-particle
energies, instead of multiples of $\hbar\Omega$,
tends to be larger for the lower-lying states
and smaller for the higher-lying states.
The single-particle energies calculated using wings tended to be 
slightly larger than those calculated 
without wings, although this difference was less than 1\%.

  The difference in energy denominators using calculated single-particle
energies results
in about a 10\% reduction in the $T=0$ and $T=1$ effective charges, 
when compared with
the traditional method of using multiples of $\hbar\Omega$, as can
be seen in Table \ref{tab4}.
Looking at the sum of the second-order diagrams in Table \ref{tab4}, we
see that the 2$\hbar\Omega$ excitation space effective charges
are much smaller, when using
calculated energies, while there is less of a difference between the two
methods for the effective charges calculated in the 4$\hbar\Omega$
and 6$\hbar\Omega$ intermediate-state excitation
spaces. In the case of the 2$\hbar\Omega$ excitation
first- and second-order diagrams, the energy denominators obtained
from calculated single-particle energies are larger and the resulting
diagrams are smaller.
All second-order diagrams have two energy denominators, which may have
calculated energy gaps that are smaller than the energy gaps determined from
multiples of $\hbar\Omega$, when we go to higher intermediate-state
excitations. For example, a configuration included in the 
6$\hbar\Omega$ excitation space, but not in the 
2$\hbar\Omega$ or 4$\hbar\Omega$ excitation
spaces, may have one of the energy denominators corresponding to a
2$\hbar\Omega$ excitation
and one corresponding to a 6$\hbar\Omega$ excitation. 
Thus the calculated energy denominator would be smaller
than 6$\hbar\Omega$ for the ``6$\hbar\Omega$" intermediate-state 
excitation and larger than 2$\hbar\Omega$ for the ``2$\hbar\Omega$" 
intermediate-state excitation. Hence, we will see
less reduction, or even an enhancement, of the diagram using
calculated single-particle energies, when compared to the
same diagram evaluated using energy denominators that are 2$\hbar\Omega$
and 6$\hbar\Omega$. This effect can be clearly seen by looking at the
2$\hbar\Omega$-excitation, 4$\hbar\Omega$-excitation,
and 6$\hbar\Omega$-excitation contributions
to the sum of the second-order diagrams, shown in Table \ref{tab4}.

  When effective-operator calculations in perturbation theory are
compared, using a $G$-matrix
computed with and without wings, there is less than a 1\% difference for the
electric quadrupole operator as well as the orbital and spin portion
of the magnetic dipole operator. Similarly, the result of using 
single-particle energies calculated from $G$-matrix elements with and 
without wings also yields less than a 1\% difference in the effective
charges obtained from the effective operators. This result holds for the 
2$\hbar\Omega$, 4$\hbar\Omega$, and 6$\hbar\Omega$ model spaces as well as
for first and second-order calculations. Since we see only a small effect 
on effective-operator results from including the wings in the calculation 
of the $G$-matrix, we can
safely conclude that the contribution of the wings plays no significant
role in our perturbation-expansion calculations.

  Because the calculations are done in a finite-sized model space, we also
investigate the effect of including more intermediate states. For this
portion of the study the same 8$\hbar\Omega$ G-matrix elements,
calculated using the Nijmegen II N-N potential, were used for all
three choices for the intermediate-state excitations. 
The same energy denominators, determined from 
calculated single-particle energies, were also employed in
all three calculations. The only difference between the three different
intermediate-state calculations is that additional excited states have been
included in the calculations. The largest intermediate-state
space includes all the configurations that are included in the 
smaller spaces and obviously yields exactly the same results 
for these lower-lying configurations. Allowing a
larger intermediate-state space will involve configurations
at  higher energies; these configurations will have
larger energy denominators and the contribution from each higher-lying
configuration should be smaller. While the contribution of each
individual configuration is generally smaller with a larger 
intermediate-excitation
space, the number of configurations greatly increases with each
increase in the excitation-space size. 

 The results of the effective-charge calculations for T=0 and T=1 
in the various model
spaces are shown in Table \ref{tab2}. In the first-order diagrams, 
only the configurations that are exactly $2\hbar\Omega$ above the 
ground state contribute. This is because the operator involved 
is an interaction 
between the $0s_{1/2}$ hole state and a particle state, and only the 
$0d_{3/2}$ and $0d_{5/2}$ particle states have non-zero matrix elements
of the transition operator
with the $0s_{1/2}$ state. Having the hole states limited to the
$0s_{1/2}$ state also excludes the possibility of a pure
$2\hbar\Omega$ excitation
configuration in the Random Phase Approximation (RPA) 
diagrams shown in Fig. \ref{1bop}(e-h).

  Because there are no additional contributions to the zeroth- and
first-order terms from excitations greater than $2\hbar\Omega$, we
are left only the second-order diagrams for investigating 
the intermediate-excitation
space dependence. In general, the contribution from the expanded portion of the 
excitation space for each individual diagram decreases as the
size of the intermediate space grows, although there are a few cases where the
$6\hbar\Omega$ excitation contribution is of the same size or slightly larger
than the $4\hbar\Omega$ excitation contribution, which is in turn of the same
size or larger than the 
$2\hbar\Omega$ excitation contribution for the same diagram.

  Siegel and Zamick \cite{SZ70} also study the
effect of going to a larger excitation space. Their method of determining
the excitation space differs from the one in this study, in that they 
restrict their space by allowing all configurations for particles within
a given shell. In the current study we restrict the intermediate 
excitation-space to all 
configurations within a given energy limit. The calculations by Siegel and
Zamick leave out
configurations with a single-particle in a high-lying shell and
a lower energy than some of their included configurations. The
calculation in the current study does not include configurations obtained 
within a given shell that are at higher energies than the limit set
on intermediate-state excitations.
Some conclusions can still be drawn from the
similarities of going to larger excitation spaces. Siegel and 
Zamick \cite{SZ70} looked at the convergence of the TDA and RPA 
diagrams and assumed that these diagrams would adequately estimate 
the effect of intermediate-space truncation. Their calculations show that the
contributions from the larger excitation space for the
TDA and RPA graphs are about one order of magnitude smaller than 
the contribution from the smaller excitation spaces. In our calculation
the contribution from each of the successively larger excitation spaces
decreases by roughly a factor of four for the TDA and RPA diagrams. 
Yet, when we calculate the sum of all second-order diagrams, we do not 
observe the same convergence that was found for the TDA and RPA studies.
In fact, the second-order total contributions from the different
intermediate-state spaces are of comparable magnitude and some are even of
different sign. This can be seen both in the results for T=0 and T=1,
and in the effective proton and neutron charges, as shown in
Table \ref{tab2}. The intermediate-state excitation-space 
truncation does not show strong
convergence, although the major contribution to the effective charges
does come from $2\hbar\Omega$ excitation contributions.

  In studying the order-by-order convergence of the diagrams, we see
again that the diagrams evaluated with T=0 
show better convergence than the diagrams evaluated with T=1.
This is just the opposite of the result found for the convergence 
of the effective interaction in Ref. \cite{HMOP96}. 
The second-order diagrams are all significantly smaller than the 
two first-order diagrams
for T=0. The sum of second-order diagrams is about an order-of-magnitude 
smaller than the sum of the two first-order diagrams. 
The second-order diagrams for T=1 are all individually smaller 
than the two first-order diagrams, but the
sum of the second-order T=1 diagrams is  approximately the same size
as the sum of the two first-order T=1 diagrams.

  When the isoscalar and isovector contributions are combined to produce
proton and neutron effective charges, via Equations (\ref{peffch})
and (\ref{neffch}),
the different convergence trends for the T=0 and T=1 diagrams cause problems
for the convergence of the effective proton and neutron 
charges. The neutron effective charge is dominated by the 
2$\hbar\Omega$ first-order
diagrams, being about twice the size of the sum of all second-order
diagrams. The proton effective charge does not show a similar dominating
term. The second-order totals show a strong excitation space dependence.
The 4$\hbar\Omega$ and 6$\hbar\Omega$ second-order totals are of
opposite sign and similar magnitude to the 2$\hbar\Omega$ second-order
total and tend to cancel any major contribution from the second-order
diagrams. It is unclear if going to larger excitation spaces will
result in any significant contributions from second-order diagrams.
Although the RPA and TDA
diagrams show good convergence as the excitation space increases, 
other second-order diagrams,
particularly ones of opposite sign, counteract this 
convergence trend. The sum of all second-order diagrams is marginally
largest for the
$2\hbar\Omega$ excitation contributions for the effective proton charge but
the same cannot be said for the effective neutron charge. 

  The effective neutron charge is of the order of 0.3 with about
two-thirds of this coming from the first-order diagram. The effective
proton charge is slightly larger than one, with the major 
contribution coming from first-order and, at least in a 6$\hbar\Omega$ 
excitation space, no significant contribution coming from
second-order diagrams.

The results for the perturbation-expansion calculation of the magnetic-dipole 
moments of $^{6}$Li are shown in Table \ref{tabdipole}. As the size
of the excitation space increases the convergence of the T=0 and T=1 
elements is well behaved for both the orbital and spin portion of the
dipole moment matrix elements. The best convergence occurs in the T=1
spin contribution to the magnetic-dipole operators, where each increase
of 2$\hbar\Omega$ in the excitation space gives an additional
contribution, which is approximately one fourth of the 
lower-excitation-space contribution.
In these particular
matrix elements the major contribution comes from the $2\hbar\Omega$
excitation
configurations with relatively small contributions from higher-lying
excitations, which tend to decrease as the excitation energy gets larger.

  The vast majority of the diagrams used to evaluate the magnetic-dipole
operator give zero contribution due to the relatively small number of
non-zero bare magnetic-dipole matrix elements.
The bare M1 matrix element is nonzero only when the
initial and final states of the operator have the same quantum numbers $n$, 
$l$. Thus, all first-order diagrams, shown in Fig. \ref{1bop},
give zero contribution to the magnetic-dipole operator because the 
operator cannot connect a $0s_{1/2}$ hole state with a higher lying
particle state. Since the zeroth-order
contribution does not change and the first-order contribution is 
identically zero for all states, it is difficult to make any statement
about the order-by-order convergence of the perturbation expansion. Even 
the TDA and RPA diagrams, which have traditionally been evaluated to higher 
order in calculations of other effective operators, are identically zero,
so we are unable to pursue the order-by-order convergence of the
magnetic-dipole operator any further.

  The results obtained from perturbation-theory calculations of the 
magnetic-dipole operator are consistent with results obtained from other 
calculations \cite{MZB66,DB69}.
All of the TDA and RPA diagrams, which we evaluate to give zero
contribution, also have zero contribution in other studies involving
one or two nucleons outside a doubly magic closed shell. The 
second-order corrections to the magnetic-dipole operator are small with the
majority contribution to these operators coming from the
zeroth-order term.

\section{MODEL-SPACE TRUNCATION RESULTS: $A$ and $\hbar\Omega$ DEPENDENCE}

For a direct comparison with the effective charges obtained in the 
per\-tur\-ba\-tion-\-ex\-pan\-sion calculation we calculate the effective 
charges in the model-space truncation scheme, as described in Ref. 
\cite{NTB97} employing the same interaction
used in the perturbation calculation, namely, a ``single-valued'' $G$-matrix
(as opposed to the effective interaction used later in this section)
with wings, obtained with the Nijmegen II N-N potential, a 
harmonic-oscillator parameter $\hbar\Omega$ = 14 MeV, and a fixed
$\Delta$ = -25 MeV. The effective charges computed with this 
interaction are shown in Table \ref{tabeffopnew} and compared with the
corresponding perturbation results in Table \ref{compare}.
These comparisons are for the one-body part of the one-plus-two-body 
effective operators obtained in this approach (see Ref. \cite{NTB97}).
The real two-body part is found to be small, typically not more than 10\%
of the full operator.

Our previous calculations of effective charges, 
published in Ref.  \cite{NTB97}, employed a different
effective two-nucleon interaction, the so-called ``multi-valued'' interaction
introduced in Ref. \cite{ZBVHS95}. In this method, different
$Q^{P}$-operators are utilized in evaluating the $G$-matrix for
configurations with different spectator energies, hence, the name
``multi-valued'' $G$-matrix approach. The effective interaction employed in
Ref. \cite{NTB97} also uses the Reid93 N-N potential
\cite{SKTdS94} and a harmonic-oscillator parameter of
$\hbar\Omega$ = 17.2 MeV.

A comparison of the effective charges from Table III of Ref. \cite{NTB97}, 
and Table \ref{tabeffopnew} in the present paper shows 
that the quadrupole effective charges 
obtained here with the ``single-valued'' $G$-matrix are considerably smaller than
those obtained with the ``multi-valued'' $G$-matrix. A major
factor in this difference is the choice of the harmonic-oscillator
parameter. To understand the dependence of the quadrupole effective 
charges on $\hbar\Omega$, we performed separate calculations,
as described later. The $\hbar\Omega$ dependence accounts for
some of the difference between the two sets of effective charges, but
the interactions used in the calculations also contribute to this
difference. The electromagnetic transitions are weaker in the
``single-valued'' calculation, which implies smaller contributions to
the effective charges. As to the comparison with the experiment,
the ``multi-valued'' interaction gives superior results. Therefore,
the effective charges extracted from the ``multi-valued'' interaction
calculation should also be more realistic.

To have a deeper insight into the dependence of the quadrupole effective 
charge on $\hbar\Omega$ and $A$, we performed several additional
calculations. 
In Fig. \ref{Ahbodep} we present the quadrupole effective proton, 
neutron (a), isoscalar (b), and isovector (c) effective charge 
dependence on $\hbar\Omega$ and $A$. 
The effective charges were obtained by the model-space truncation method, 
going from the $6\hbar\Omega$ relative to the unperturbed ground state
to the $0\hbar\Omega$ model space  
in calculations for $^6$Li and $^5$He, $^5$Li. 
The $A=5$ results were extracted
from the calculations presented in Fig.1 of Ref. \protect\cite{NB96}.
The calculations were performed for a wide range of the harmonic-oscillator
frequencies, $\hbar\Omega=8,10,14,17.8$, and 22 MeV.
The $A=6$ results obtained for $\hbar\Omega=17.2$ MeV are taken from 
Ref. \protect\cite{NTB97}. Two more 
$A=6$ calculations, for $\hbar\Omega=14$ and 20 MeV, were performed,
using the same approach.
All the calculations use the ``multi-valued'' isopin-invariant
effective interaction derived from the Reid93 potential, as described
in Ref. \protect\cite{NB96}. The effective charges are trivially computed
for the $A=5$ system by taking the ratios of the matrix elements
calculated in the $6\hbar\Omega$ space with the corresponding ones
from the $0\hbar\Omega$ space, e.g., 
$e_{\rm eff33}^{\rm p}=
\frac{\langle\frac{3}{2}^-(6\hbar\Omega,^5{\rm Li})|
Q^{(2)}|\frac{3}{2}^-(6\hbar\Omega,^5{\rm Li})\rangle}    
{\langle\frac{3}{2}^-(0\hbar\Omega,^5{\rm Li})|
Q^{(2)}|\frac{3}{2}^-(0\hbar\Omega,^5{\rm Li})\rangle}$. The one-body
quadrupole operator used in the calculations employs free nucleon 
charges.
The $A=6$ effective charges are obtained using the method described
in Ref. \cite{NTB97}.

 From Fig. \ref{Ahbodep} we observe an almost linear scaling of the 
effective charges. This can be simply understood. In particular, for
the $A=5$ results, it is apparent that such scaling exists under the 
condition that the large-space, here $6\hbar\Omega$, results depend 
only weakly on $\hbar\Omega$. The $0\hbar\Omega$ results, 
on the other hand, are proportional to the harmonic-oscillator
parameter $b^2 = (\hbar c)^2/(m_{\rm N}c^2\hbar\Omega)$.
Therefore, $e_{\rm eff}(\hbar\Omega_1)\approx \frac{\hbar\Omega_1}
{\hbar\Omega_2}e_{\rm eff}(\hbar\Omega_2)$. Apparently, and not
surprisingly, the scaling persists for $A=6$ as well.

A more non-trivial result is, however, the observation that the isoscalar
effective charges remain almost the same for $A=5$ and $A=6$, while,
on the other hand, there is a significant change in the isovector
effective charges. This can be seen by comparing 
parts (b) and (c) of Fig. \ref{Ahbodep}.
In fact, a similar $A$ dependence was reported by Nakada and Otsuka in 
Ref. \cite{NaOt94} in a phenomenological shell-model calculation
for the p-shell nuclei. Here we make a similar observation 
in a microscopic calculation.
 
Let us return to the comparison of the effective charges obtained from the
``single-valued'' effective interaction, Table \ref{tabeffopnew},
and from the ``multi-valued'' interaction, Table III of Ref. \cite{NTB97}. From 
the above discussion of the  $\hbar\Omega$ dependence,
we should expect the electric-quadrupole effective charges,
calculated with $\hbar\Omega$ = 17.2 MeV, to be larger by a factor of
$17.2/14 \approx 1.23$. The remaining difference in the effective charges
should then be attributed to the differences in the two effective interactions.

\section{ CONCLUSIONS }

  To summarize, the goal of this project was to study different
microscopical approaches for calculating the electromagnetic
operators of light nuclei, and in particular, $^{6}$Li.
As in Ref. \cite{TZB96} for the effective interaction,  
the no-core calculation was again used 
as a ``theoretical experiment'' and as a starting point for the model-space
truncation calculation for the purpose of comparison with the
standard perturbation-theory calculations of effective operators. The
results for the electric quadrupole operator give us more information
about perturbation-theory calculations of effective operators than the
results of the magnetic dipole calculation. For the electric
quadrupole operator we find that the use of single-particle energies,
instead of multiples of $\hbar\Omega$, for the energy denominators has
a small effect, favoring larger effective proton and
neutron charges for calculations
with energy denominators, which are multiples of $\hbar\Omega$. Hence, 
in perturbation-theory calculations of effective charges, calculated 
single-particle energies appear to have no advantage over multiples of
$\hbar\Omega$.

  Similar to previous perturbation-theory calculations of effective charges
\cite{EO77,SZ70,ES71}, we find that the perturbation expansion does not
converge rapidly.  In addition, we discover, as seen in Tables III and IV,
that the second-order terms are not well-behaved as the size of the space
increases.  This is opposite to the conclusion of Siegel and Zamick 
\cite{SZ70}, based only on the RPA and TDA terms (which are also well-behaved
with space size in our calculations).  It is unclear whether going to 
larger excitation spaces and/or higher-orders in the perturbation expansion
will change the conclusions based on the present results.
 
  Unlike the calculation for the effective interaction,
the order-by-order convergence of the perturbation expansion for
effective charges is better for $T=0$ states than for $T=1$ states
\cite{TZB96,HMOP96}.
When we look at the proton and neutron effective charges computed 
in the perturbation theory, we see that
the proton effective charge has small and opposite contributions
coming from the first- and second-order terms, while the effective
neutron charge has its first-order contribution about twice as large
and of the same sign as the second-order contribution. 
The effective
charges that we obtain, e$_{n}\approx$ 0.3 and e$_{p}\approx$ 1.1, are
similar to the effective charges obtained 
in other perturbation calculations for larger nuclei. The
effective charges obtained through a perturbation expansion
show good agreement with the effective charges obtained through
a model-space truncation calculation, when the same interaction
and harmonic-oscillator parameter are used.

  Since the zeroth-order effective charges are fixed and the
first-order contributions are zero for the magnetic dipole operator,
there is little that can be concluded about the convergence of this
operator. It is worth noting that there is very little contribution to
both the orbital and the spin portion of the magnetic-dipole
operator from second-order. This is consistent with the findings of
the model-space-truncation results for effective magnetic-dipole 
charges, i.e. g-factors. Similar to the findings from the
calculation of the electric-quadrupole operator,
the effective charges we obtain from
the perturbation expansion for the magnetic dipole operator are of the
same size as effective charges obtained in the model-space-truncation 
calculation, when using the same ``single-valued'' $G$-matrix in both
calculations, as summarized in Table \ref{compare}.

  We have also studied the $\hbar\Omega$ and the $A$ dependence of the 
E2 effective charges in the model-space-truncation method.
We used the $6\hbar\Omega$ calculations for $^5$Li, $^5$He, and $^6$Li,
performed for a wide range of the harmonic-oscillator frequencies and 
using the ``multi-valued'' effective interaction derived from the Reid93
potential, to extract the $0\hbar\Omega$ effective charges. We find
a scaling of the effective charges with $\hbar\Omega$, which can be simply
understood. 
Let us mention that the effective charges extracted from the experimental
E2 transitions or moments depend on the value of $\hbar\Omega$ used in the 
analysis and would scale with $\hbar\Omega$ in a similar manner as we 
observe in our microscopic calculations. Typically, the appropriate values 
of $\hbar\Omega$ employed in such analyses were taken from a standard formula,
such as, $\hbar\Omega=45 A^{-\frac{1}{3}}-25 A^{-\frac{2}{3}}$ MeV.
For $A=6$ this gives 17.2 MeV, the same value we utilized in our calculations
in Ref. \cite{NTB97},
and for which we obtained effective charges consistent with the experimental
ones.   
Also, we observe different behavior of the isoscalar
and isovector effective charges with respect to a change in the mass 
number $A$. While the isoscalar charges remain almost constant as a function
of $A$, the isovector charges change significantly between $A=5$ and $A=6$.
A similar observation for other p-shell nuclei was reported in
Ref. \cite{NaOt94}. There is also a dependence of the effective 
charge on the strength of the effective interaction used 
in the calculation. We obtain the significant result that the 
``multi-valued'' effective interaction yields more realistic values for 
the effective charges than the ``singled-valued'' effective interaction.

\acknowledgements{
This material is based upon work supported by the 
National Science Foundation under Grant No. PHY96-05192.
P.N. also acknowledges partial support from
the grant of the Grant Agency of the Czech Republic 202/96/1562.
}

\begin{figure} [h]
\centering
\caption{$Q^{P}$-space projection operator for a 6$\hbar\Omega$
calculation of $^{6}$Li (a)with wings and (b)without wings. }
\label{wingsfig}
\end{figure}

\begin{figure}[h]
\centering
\caption{(a)Zeroth- and (b),(c), first-order one-body
diagrams for the effective charges. (d-f) illustrate some 
second-order,
(g-h) the TDA and (i-l) the RPA diagrams included in the calculation
of the effective charges.}
\label{1bop}
\end{figure}

\begin{figure}[h]
\centering
\caption{First- and second-order
one-body diagrams used to evaluate the single-particle energies.}
\label{spe2nd}
\end{figure}

\begin{figure}[h]
\caption{(a) The quadrupole effective proton, neutron, (b) isoscalar, 
and (c) isovector effective charge dependence on $\hbar\Omega$ and $A$. 
The effective charges were obtained by the model-space truncation 
from $6\hbar\Omega$
calculations for $^6$Li and $^5$He, $^5$Li. The $A=5$ results correspond
to the calculation presented in Fig.1 of Ref. \protect\cite{NB96},
obtained for $\hbar\Omega=8,10,14,17.8$, and 22 MeV.
The $A=6$ results, obtained for $\hbar\Omega=17.2$ MeV, are taken from 
Ref. \protect\cite{NTB97}. 
Two additional calculations, for $\hbar\Omega=14$ and 20 MeV, are
presented.
All the calculations use the ``multi-valued''
effective interaction derived from the Reid93 potential, as described
in Ref. \protect\cite{NB96}. 
}
\label{Ahbodep}
\end{figure}

\begin{table}[h]
\centering
\begin{tabular}{|l||c|c||c|c||c|c||c|}
 & \multicolumn{2}{c||}{2$\hbar\Omega$} & \multicolumn{2}{c||}
{4$\hbar\Omega$} & \multicolumn{2}{c||}{6$\hbar\Omega$} & Experiment \\
Operator & wings &no wing& wings &no wing& wings &no wing& \\
\hline
Gs.E. [MeV]   &-24.18 &-26.82 &-31.61 &-32.23 &-31.92 &-32.07&-32.00 \\
E2 [$e$ fm$^2$]   &-0.039& -0.108& -0.075& -0.109& -0.089& -0.092& -0.082 \\
M1 [$\mu_{\rm N}$] & 0.861 & 0.860 & 0.850 & 0.850 & 0.844 & 0.844& 0.822 \\
Rp [fm] & 2.072 & 2.028 & 2.061 & 2.046 & 2.075 & 2.069& 2.38 \\
\end{tabular}
\caption{No-core results for the ground-state energy (Gs.E.),
electric quadrupole moment (E2),
magnetic di\-pole moment (M1), and the root-mean-square point-proton
radius (Rp) calculated
in different model-space sizes for a $G$-matrix calculated
with and without wings on the $Q^{P}$-operator.}
\label{nocoropwing}
\end{table}

\begin{table} [h]
\centering
\begin{tabular}{|l||cc|cc|}
2, 4, and 6$\hbar\Omega$ excitation contributions to
& \multicolumn{2}{c|}{multiples of $\hbar\Omega$}
& \multicolumn{2}{c|} {Calculated s.p.e.} \\
first- and second-order totals of $0p_{1/2}-0p_{3/2}$& T=0 & T=1 & T=0 & T=1 \\
\hline
\hline
$2\hbar\Omega$ excitations first-order & 0.3270 &-0.1493 & 0.2706 & -0.1213 \\
\hline
$2\hbar\Omega$ excitations second-order &-0.0992&-0.1289&-0.0288&-0.0712 \\
$4\hbar\Omega$ excitations second-order & 0.1171&-0.0244& 0.0913&-0.0226 \\
$6\hbar\Omega$ excitations second-order & 0.0528&-0.0226& 0.0530&-0.0218 \\
\hline
\multicolumn{5}{|l|}{Total $1^{st}$ and $2^{nd}$ order for all configurations
within a given model space} \\
\hline
$2\hbar\Omega$ excitations $1^{st}$+$2^{nd}$ total
& 0.2278 &-0.2782 & 0.2418 &-0.1925 \\
$(2+4)\hbar\Omega$ excitations $1^{st}$+$2^{nd}$ total
& 0.3449 &-0.3026 & 0.3331 &-0.2151 \\
$(2+4+6)\hbar\Omega$ excitations $1^{st}$+$2^{nd}$ total
& 0.3977 &-0.3252 & 0.3861 &-0.2369 \\
\hline
\hline
$0_{th}+1^{st}+2^{nd}$ order & proton & neutron & proton & neutron \\
$(2+4+6)\hbar\Omega$ excitations & 1.0363 & 0.3615 & 1.0746 & 0.3115 \\
\end{tabular}
\caption{Results of perturbation-expansion calculations, showing the dependence
of effective charges upon the type of energy denominators used. Results
shown are for the $0p_{1/2}-0p_{3/2}$ transitions calculated without
wave function renormalization diagrams. In the last two columns calculated
single-particle energies (s.p.e.) are used (see Fig. 3).}
\label{tab4}
\end{table}

\begin{table} [h]
\centering
\begin{tabular}{l||cc|cc||cc|cc}
 & \multicolumn{4}{c|} { $0p_{1/2} - 0p_{3/2}$ }
 & \multicolumn{4}{|c} { $0p_{3/2} - 0p_{3/2}$ } \\
 & T = 0& T = 1& proton& neutron&T = 0 &T = 1 &proton &neutron \\
\hline
\hline
$2\hbar\Omega$ excitations $1^{st}$ order
& 0.2706&-0.1213& 0.0747& 0.1960& 0.2257&-0.1328& 0.0464& 0.1793\\
$4\hbar\Omega$ excitations $1^{st}$ order 
& --- & --- & --- & --- & --- & --- & --- & --- \\
$6\hbar\Omega$ excitations $1^{st}$ order
& --- & --- & --- & --- & --- & --- & --- & --- \\
\hline
$2\hbar\Omega$ excitations $2^{nd}$ order TDA
& 0.0517& 0.0115& 0.0316& 0.0201& 0.0478& 0.0122& 0.0300& 0.0178\\
$4\hbar\Omega$ excitations $2^{nd}$ order TDA
& 0.0158& 0.0040& 0.0099& 0.0059& 0.0146& 0.0043& 0.0095& 0.0052\\
$6\hbar\Omega$ excitations $2^{nd}$ order TDA
& 0.0036& 0.0011& 0.0023& 0.0013& 0.0033& 0.0011& 0.0022& 0.0011\\
\hline
$2\hbar\Omega$ excitations $2^{nd}$ order RPA 
& --- & --- & --- & --- & --- & --- & --- & --- \\
$4\hbar\Omega$ excitations $2^{nd}$ order RPA
& 0.0226& 0.0048& 0.0137& 0.0089& 0.0191& 0.0053& 0.0122& 0.0069\\
$6\hbar\Omega$ excitations $2^{nd}$ order RPA
& 0.0068& 0.0017& 0.0043& 0.0025& 0.0056& 0.0020& 0.0038& 0.0018\\
\hline
$2\hbar\Omega$ excitations $2^{nd}$ order total
&-0.0289&-0.0712&-0.0500& 0.0212&-0.0346&-0.0630&-0.0488& 0.0142\\
$4\hbar\Omega$ excitations $2^{nd}$ order total
& 0.0913&-0.0226& 0.0344& 0.0570& 0.0854&-0.0212& 0.0321& 0.0533\\
$6\hbar\Omega$ excitations $2^{nd}$ order total
& 0.0530&-0.0218& 0.0156& 0.0374& 0.0506&-0.0211& 0.0148& 0.0358\\
\hline
$2\hbar\Omega$ excitations $1^{st}$+$2^{nd}$ total
& 0.2418&-0.1924& 0.0247& 0.2171& 0.1911&-0.1958&-0.0024& 0.1935\\
$4\hbar\Omega$ excitations $1^{st}$+$2^{nd}$ total
& 0.0913&-0.0226& 0.0344& 0.0570& 0.0854&-0.0212& 0.0321& 0.0533\\
$6\hbar\Omega$ excitations $1^{st}$+$2^{nd}$ total
& 0.0530&-0.0218& 0.0156& 0.0374& 0.0506&-0.0211& 0.0148& 0.0358\\
\hline
$0^{th}$+$1^{st}$+$2^{nd}$ order total &1.3861 &0.7632 
&1.0747 & 0.3115 &1.3271 &0.7619 & 1.0445 & 0.2826 \\
\end{tabular}
\caption{Iso-scalar and Iso-vector components as well as
proton and neutron quadrupole effective charges
for various diagrams using calculated single-particle energies and
an 8$\hbar\Omega$ $G$-matrix with wings.
Numbers listed are contributions due to the
expanded portion of the intermediate-excitation space only, calculated
without wave function renormalization diagrams.}
\label{tab2}
\end{table}

\begin{table} [h]
\centering
\begin{tabular}{|l|c|c|c|c|c|l|}
Excitation & \multicolumn{2}{c|} { $0p_{1/2} - 0p_{1/2}$ }
 & \multicolumn{2}{c|} { $0p_{1/2} - 0p_{3/2}$ }
 & \multicolumn{2}{c|} { $0p_{3/2} - 0p_{3/2}$ } \\
Space & T = 0 & T = 1 &  T = 0 &  T = 1 & T = 0 & T = 1 \\
\hline
2 $\hbar\Omega$ l-part &-0.0765& 0.0243&-0.0135& 0.0147& 0.0279& 0.0213\\
4 $\hbar\Omega$ l-part &-0.1020& 0.0410&-0.0220& 0.0253& 0.0307& 0.0361\\
6 $\hbar\Omega$ l-part &-0.1146& 0.0527&-0.0296& 0.0310& 0.0267& 0.0432\\
\hline
2 $\hbar\Omega$ s-part & 0.0174&-0.0439&-0.0135&-0.0364& 0.0375&-0.0228\\
4 $\hbar\Omega$ s-part & 0.0329&-0.0537&-0.0220&-0.0452& 0.0753&-0.0286\\
6 $\hbar\Omega$ s-part & 0.0400&-0.0560&-0.0319&-0.0481& 0.0943&-0.0304\\
\hline
\hline
 & \multicolumn{2}{c|} { $0p_{1/2} - 0p_{1/2}$ }
 & \multicolumn{2}{c|} { $0p_{1/2} - 0p_{3/2}$ }
 & \multicolumn{2}{c|} { $0p_{3/2} - 0p_{3/2}$ } \\
 & proton & neutron & proton & neutron & proton & neutron \\
\hline
6 $\hbar\Omega$ l-part&-0.0310&-0.0837&0.0007&-0.0303&0.0350&-0.0083\\
6 $\hbar\Omega$ s-part&-0.0080& 0.0480&-0.0400& 0.0081&0.0320& 0.0623\\
\end{tabular}
\caption{Second order terms in the perturbation-expansion 
calculation of the magnetic
dipole moment of $^{6}$Li using calculated single-particle energies and
an 8$\hbar\Omega$ $G$-matrix with wings.
$N\hbar\Omega$ means the sum of all contributions from 
2 to $N\hbar\Omega$.
}
\label{tabdipole}
\end{table}

\begin{table} [h]
\centering
\begin{tabular}{|c|rrrrrr|}
 & $e^{\rm p}_{\frac{1}{2}\frac{3}{2}}$
 & $e^{\rm n}_{\frac{1}{2}\frac{3}{2}}$
 & $e^{\rm p}_{\frac{3}{2}\frac{3}{2}}$
 & $e^{\rm n}_{\frac{3}{2}\frac{3}{2}}$
 & $e^{\rm p}_{\frac{1}{2}\frac{1}{2}}$
 & $e^{\rm n}_{\frac{1}{2}\frac{1}{2}}$ \\
\hline
E2 & 1.184 & 0.272 & 1.080 & 0.215 & - & - \\
Ml & 0.890 & 0.036 & 0.954 & 0.061 & 0.910 & 0.068 \\
Ms & 0.939 &-0.012 & 0.972 &-0.001 & 0.943 &-0.031 \\
\hline
 & $e^{\rm p}_{\rm eff}$ & $e^{\rm n}_{\rm eff}$
 & $e^{\rm p}_{\rm eff-4}$ & $e^{\rm n}_{\rm eff-4}$ & & \\
\hline
E2 & 1.141 & 0.272 & 1.087 & 0.189 & & \\
Ml & 0.934 & 0.061 & 0.948 & 0.047 & & \\
Ms & 0.955 &-0.005 & 0.966 &-0.005 & & \\
\end{tabular}
\caption{Effective charges of the proton and neutron quadrupole,
magnetic orbital and magnetic spin operators, derived by least-square
fits to the corresponding $0p$-shell effective operators obtained 
by model-space truncation method from
the 6$\hbar\Omega$ calculation for $^6$Li using the 
``single-valued" $G$-matrix with wings, $\Delta$ = -25 MeV, 
and $\hbar\Omega$ = 14 MeV. Both the $j$-dependent
and $j$-independent effective charges are shown.
Also, the $j$-independent effective charges obtained in the same way
from the 4$\hbar\Omega$ calculation are presented in the last two
columns, labelled eff-4.}
\label{tabeffopnew}
\end{table}

\begin{table} [h]
\centering
\begin{tabular}{|l||c|c||c|c||c|c|}
 & \multicolumn{2}{c||} { $0p_{1/2} - 0p_{1/2}$ }
 & \multicolumn{2}{c||} { $0p_{1/2} - 0p_{3/2}$ }
 & \multicolumn{2}{c|} { $0p_{3/2} - 0p_{3/2}$ } \\
 & proton & neutron & proton & neutron & proton & neutron \\
\hline
E2 perturbation & --- & --- & 1.07 & 0.31 & 1.04 & 0.28 \\
E2 space truncation & --- & --- & 1.18 & 0.27 & 1.08 & 0.22 \\
\hline
M1l perturbation & 0.97 & -0.08 & 1.00 & -0.03 & 1.04 & -0.01 \\
M1l space truncation & 0.91 & 0.07 & 0.89 & 0.04 & 0.95 & 0.06 \\
\hline
M1s perturbation & 0.99 & 0.05 & 0.96 & 0.01 & 1.03 & 0.06 \\
M1s space truncation & 0.94 & -0.03 & 0.94 & -0.01 & 0.97 &-0.001 \\
\end{tabular}
\caption{Comparison of effective charges obtained through a
per\-tur\-ba\-tion-ex\-pan\-sion calculation and a model-space-truncation
calculation. Both results are for a ``single-valued,'' 8$\hbar\Omega$
$G$-matrix with wings derived using the Nijmegen II N-N 
potential, $\Delta$ = -25 MeV, and $\hbar\Omega$ = 14 MeV.}
\label{compare}
\end{table}

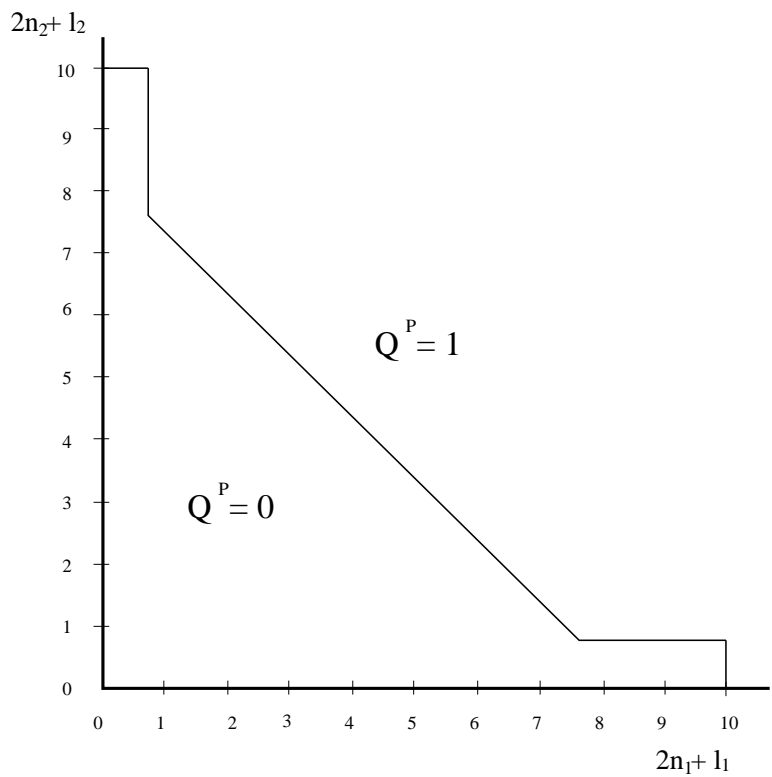
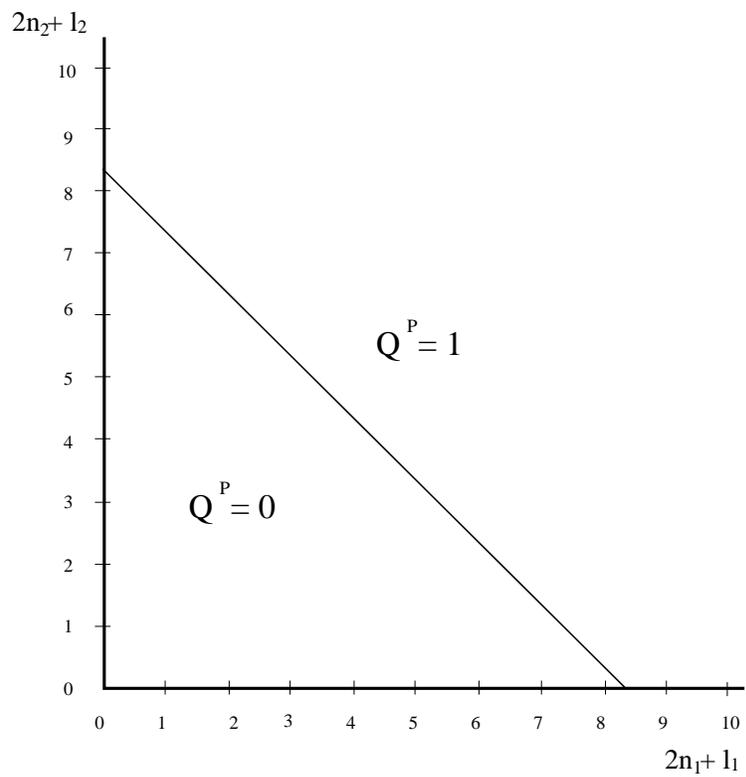
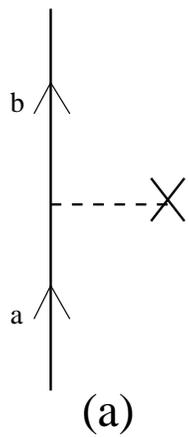
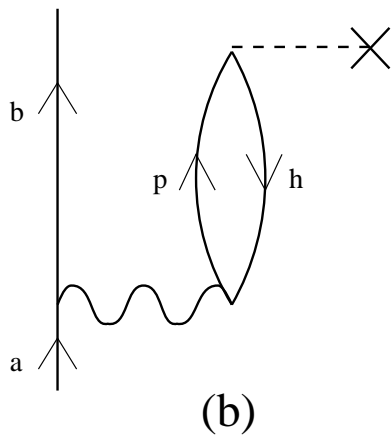
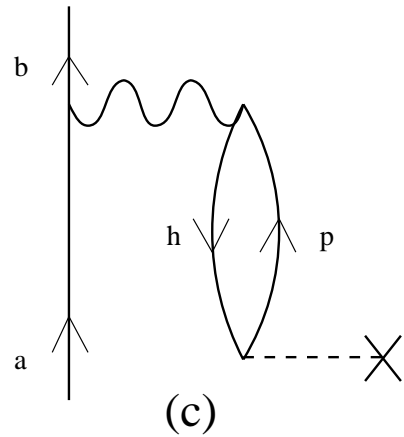
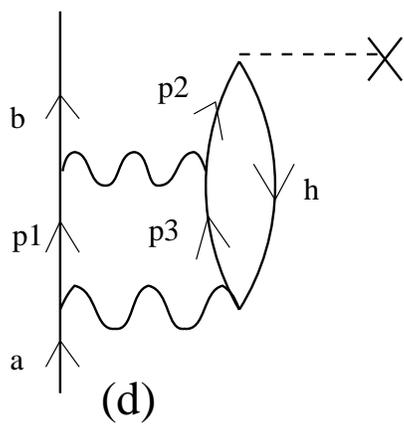
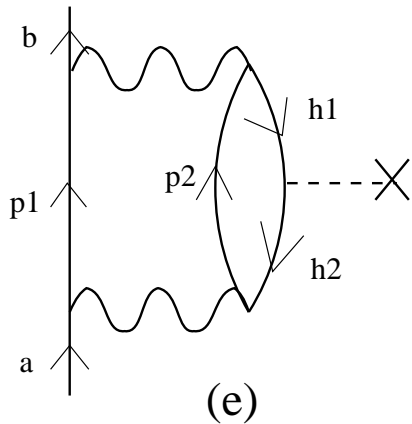
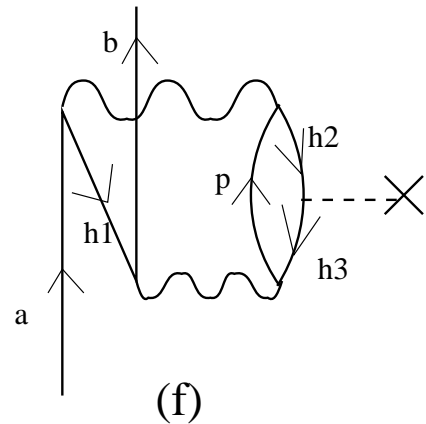
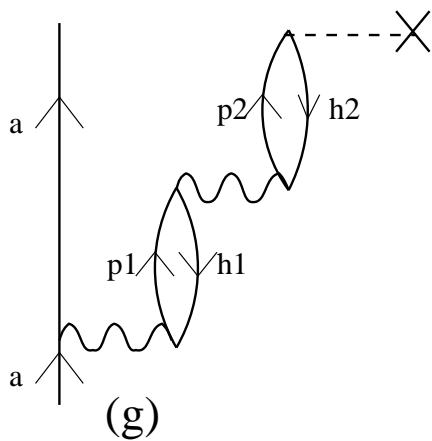
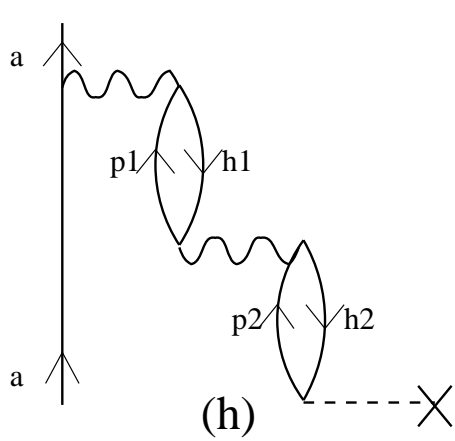
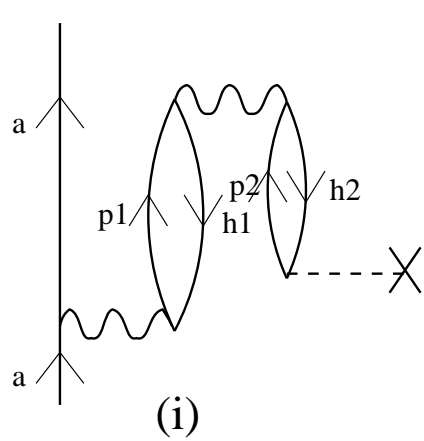
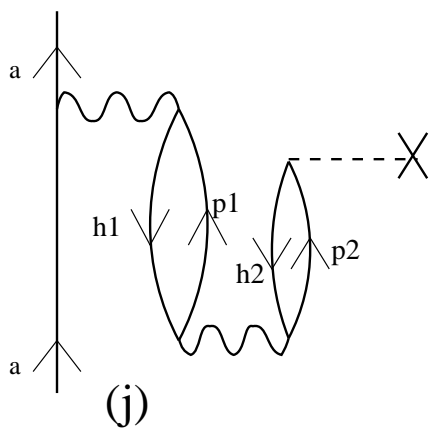
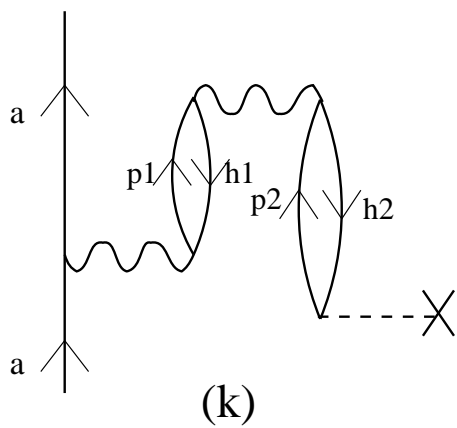
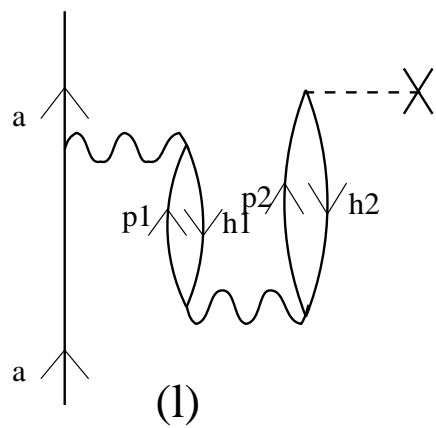
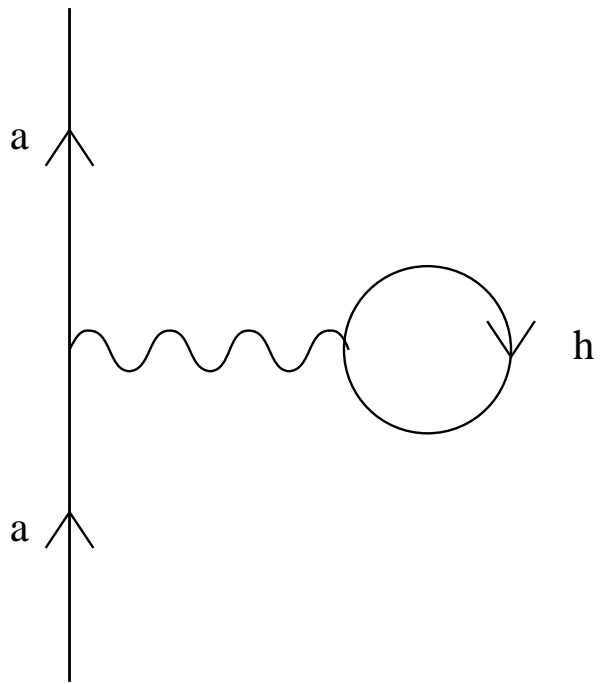
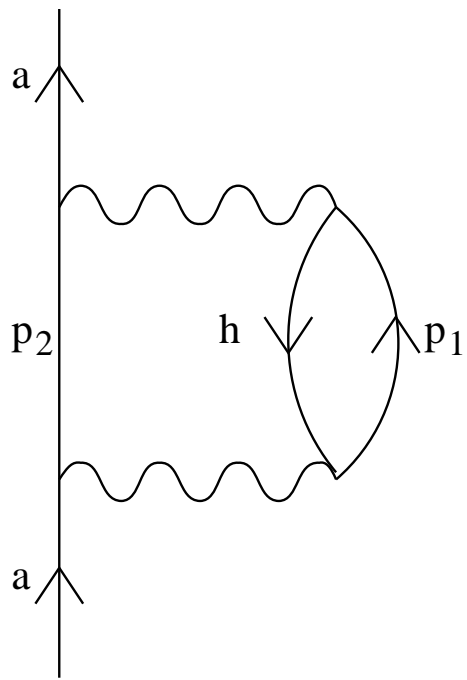
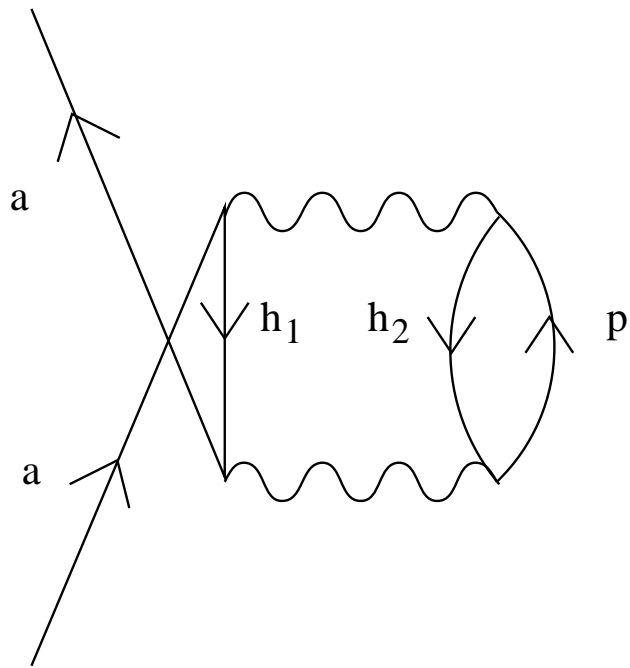
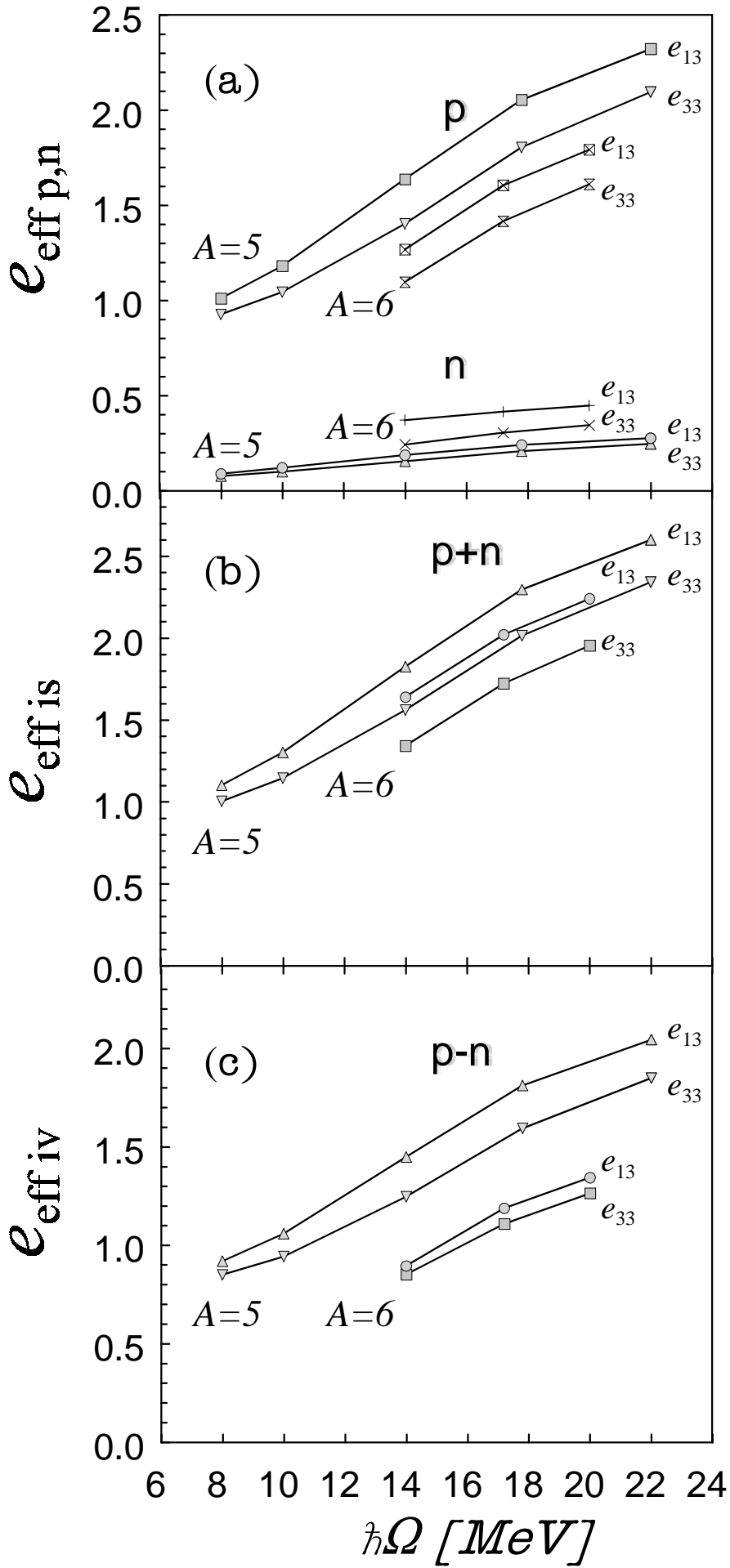
\end{document}